\documentclass[11pt]{article}
\usepackage{epsfig}
\usepackage{graphicx}

\newcommand{\BABARPubYear}    {00}

\newcommand{\BABARProcNumber} {20}
\newcommand{\SLACPubNumber} {8733}

\usepackage{relsize}
\def\babar{\mbox{\slshape B\kern-0.1em{\smaller A}\kern-0.1em
    B\kern-0.1em{\smaller A\kern-0.2em R}}}

\def\Kbar  {\kern 0.2em\overline{\kern -0.2em K}{}}

\def\Kzb   {\ensuremath{\Kbar^0}}
\def\KzKzb {\ensuremath{K^0 \kern -0.16em \Kzb}}
\def\Dz    {\ensuremath{D^0}}
\def\Dbar  {\kern 0.2em\overline{\kern -0.2em D}{}}

\def\Dzb   {\ensuremath{\Dbar^0}}
\def\DzDzb {\ensuremath{D^0 {\kern -0.16em \Dzb}}}
\def\Dstar   {\ensuremath{D^*}}

\def\Bz    {\ensuremath{B^0}}

\def\Bbar  {\kern 0.18em\overline{\kern -0.18em B}{}}

\def\Bzb   {\ensuremath{\Bbar^0}}

\def\BzBzb {\ensuremath{B^0 {\kern -0.16em \Bzb}}}

\mathchardef\Upsilon="7107
\def\Y#1S{\ensuremath{\Upsilon{(#1S)}}}

\mathchardef\Deltares="7101
\mathchardef\Xi="7104
\mathchardef\Lambda="7103
\mathchardef\Sigma="7106
\mathchardef\Omega="710A
\def\Deltabar   {\kern 0.25em\overline{\kern -0.25em \Deltares}{}}
\def\Lbar {\kern 0.2em\overline{\kern -0.2em\Lambda\kern 0.05em}\kern-0.05em{}}
\def\Sigbar{\kern 0.2em\overline{\kern -0.2em \Sigma}{}}
\def\Xibar{\kern 0.2em\overline{\kern -0.2em \Xi}{}}
\def\Obar{\kern 0.2em\overline{\kern -0.2em \Omega}{}}
\def\Nbar{\kern 0.2em\overline{\kern -0.2em N}{}}
\def\Xbar{\kern 0.2em\overline{\kern -0.2em X}{}}

\def\upsbb {\ensuremath{\Upsilon{\rm( 4S)}\to B\Bbar}}

\def\ev   {\ensuremath{\rm \,e\kern -0.08em V}}
\def\kev  {\ensuremath{\rm \,ke\kern -0.08em V}} 
\def\mev  {\ensuremath{\rm \,Me\kern -0.08em V}} 
\def\gev  {\ensuremath{\rm \,Ge\kern -0.08em V}} 
\def\gevc {\ensuremath{{\rm \,Ge\kern -0.08em V\!/}c}} 
\def\tev  {\ensuremath{\rm \,Te\kern -0.08em V}}
\def\mevc {\ensuremath{{\rm \,Me\kern -0.08em V\!/}c}} 
\def\gevcc{\ensuremath{{\rm \,Ge\kern -0.08em V\!/}c^2}} 
\def\mevcc{\ensuremath{{\rm \,Me\kern -0.08em V\!/}c^2}}

\def\mus  {\ensuremath{\rm \,\mus}}

\def\mus        {\ensuremath{\,\mu{\rm s}}}    
\def\gsim{{~\raise.15em\hbox{$>$}\kern-.85em
          \lower.35em\hbox{$\sim$}~}}
\def\lsim{{~\raise.15em\hbox{$<$}\kern-.85em
          \lower.35em\hbox{$\sim$}~}}

\def\to                 {\ensuremath{\rightarrow}}

\def\pep2{PEP-II}

\providecommand{\eqref}[1]{Eq.~(\ref{eq:#1})}

\newcommand{\epjc}      [1]  {{Eur.\ Phys.\ Jour.\ C~{\bf #1}}}

\providecommand{\pl}        [1]  {{Phys.\ Lett.\ {\bf #1}}}      

\newcommand{\zp}        [1]  {{Z.\ Phys.\ {\bf #1}}}

\def\jetset74   {\mbox{\tt Jetset \hspace{-0.5em}7.\hspace{-0.2em}4}}

\newcommand{\nonumsection}[1] {\vspace{12pt}\noindent{#1}
	\par\vspace{5pt}}
\newcommand{\textlineskip}{\baselineskip=13pt}
\def\figurebox#1#2#3{%
        \def\arg{#3}%
        \ifx\arg\empty
        {\hfill\vbox{\hsize#2\hrule\hbox to #2{\vrule\hfill\vbox to #1{\hsize#2\vfill}\vrule}\hrule}\hfill}%
        \else
        {\hfill\epsfbox{#3}\hfill}%
        \fi}

\long\def\inst#1{\par\nobreak\kern 4pt\nobreak
    {\it #1}\par\vskip 10pt plus 3pt minus 3pt}

\begin{document}
{\pagestyle{empty}

\begin{flushright}
SLAC-PUB-\SLACPubNumber \\
\babar-PROC-\BABARPubYear/\BABARProcNumber \\
December, 2000 \\
\end{flushright}

\par\vskip 3cm

\begin{center}
\Large \bf $B$ DECAYS TO FINAL STATES INCLUDING $D_s^{(*)}$ AND $D^*$
\end{center}
\bigskip

\begin{center}
\large 
Benjamin Brau\\
bbrau@slac.stanford.edu\\
(representing the \babar\ collaboration)\\[5pt]
Laboratory for Nuclear Science\\
Massachusetts Institute of Technology\\
77 Massachusetts Avenue\\
Cambridge MA 02139, USA
\end{center}
\bigskip \bigskip

\begin{center}
\large \bf Abstract
\end{center}
The $e^+e^-$ annihilation data recorded with the \babar\ detector has
been used to study $B$ decays to $D_s^{(*)\pm}$ and $D^{*\pm}$ mesons. 
The production fraction of 
inclusive $D_s^{(*)\pm}$ and the corresponding momentum spectra have
been determined.
Exclusive decays $B^0 \rightarrow D^{*-}D_s^{(*)+}$ have been identified
with a partial reconstruction technique and their branching fractions have
been measured. 
We also report branching fraction measurements for the exclusive hadronic
modes $B^0 \rightarrow D^{*-} \pi^+$ and $B^0 \rightarrow D^{*-} \rho^+$

\vfill
\begin{center}
Contribued to the
Proceedings of the\\
Meeting of the Division of Particles
and Fields of the American Physical Society,\\ 
8/9/2000---8/12/2000, Columbus, Ohio, USA
\end{center}

\vspace{1.0cm}
\begin{center}
{\em Stanford Linear Accelerator Center, Stanford University, 
Stanford, CA 94309} \\ \vspace{0.1cm}\hrule\vspace{0.1cm}
Work supported in part by Department of Energy contract DE-AC03-76SF00515.
\end{center}

\vspace*{1pt}\textlineskip	
\section{Introduction}	
\vspace*{-0.5pt}
\noindent

The study of $D_s^{(*)}$ production in $B$ decays allows us to understand
the mechanisms leading to the creation of $c\bar{s}$ quark pairs.
The precise measurement of the momentum spectrum allows a determination of
the fraction of two-body and multi-body decay modes, and
helps in understanding $b \rightarrow c\bar{c}s$ transitions.
In this study we report a new measurement of the inclusive
$D_s^{(*)\pm}$ production rate in $B$ decays and
the branching fractions of two specific two-body $B$ decay modes
involving a $D_s^{(*)\pm}$ meson. 
We have also studied the decay modes $B^0 \rightarrow
D^{*-} \pi^+$ and $B^0 \rightarrow D^{*-} \rho^+$ and measured the
corresponding branching fractions using fully-reconstructed events.
Those measurements are interesting for testing factorization models of B decays to open charm.

\section{Dataset}

\noindent

The data were collected with the \babar\ detector
operating at the PEP-II storage ring at the Stanford Linear
Accelerator Center.
 For the inclusive $D_s^{(*)}$ production in $B$ decays and the 
$B^0 \rightarrow D^{*-} D_s^{(*)+}$ branching fraction measurements we used a data
sample equivalent to $7.7$ fb$^{-1}$ of integrated luminosity collected
while running on the $\upsbb$ resonance and a sample of $1.2$ fb$^{-1}$ 
collected $40$ \mevcc\ below the $B\bar{B}$ threshold.
Branching fraction measurements for the $B^0 \rightarrow D^{*-}
\pi^+$ and $B^0 \rightarrow D^{*-} \rho^+$ modes use  a
subset corresponding to an integrated luminosity 
of $5.2$ fb$^{-1}$.

\section{Inclusive $D_s^{(*)}$ Production in $B$ Decay}
\noindent

The $D_s^{\pm}$ mesons are reconstructed in the $D_s^{\pm} \rightarrow
\phi \pi^{\pm}$ decay mode with $\phi \rightarrow K^+K^-$.
Particle identification is used to obtain a clean sample. 
Three charged tracks coming from a common vertex are combined to form a 
$D_s^{\pm}$. Two of them, with opposite charge, are required to be
positively identified  as
kaons and their invariant mass must be within $8$ \mev\ of the nominal
$\phi$ mass. In this decay the $\phi$ meson is polarized longitudinally
which means the helicity angle of the decay, $\theta_H$  has a  $\cos^2
\theta_H$ dependence\cite{conf0013}. The requirement $|\cos
\theta_H|>0.3$ keeps $97.5\%$ of the signal while
rejecting $30 \%$ of the background.
Candidate $D_s^{*\pm}$ mesons are reconstructed in the decay channel $D_s^{*\pm}
\rightarrow D_s^{\pm} \gamma$ with $D_s^{\pm} \rightarrow \phi
\pi^{\pm}$. $D_s^{\pm}$ candidates are required to be within $2.5\sigma$
of the $\phi\pi$ invariant mass. Photon candidates must deposit a minimum energy
of $50$ \mevcc\ in the calorimeter.
The number of $D_s^{\pm}$ mesons is extracted from a Gaussian fit of the 
$\phi\pi^{\pm}$ invariant mass distribution for different momentum
ranges in the $\upsbb$ rest frame.
Similarly, the number of $D_s^{*\pm}$ is extracted by fitting
the mass difference $m_{D_s^{*}}-m_{D_s}$ distribution.    
The efficiency-corrected number of reconstructed $D_s^{(*)\pm}$ as
a function of momentum is shown in
Fig.~\ref{fig:momspectra}.
To determine the $D_s^{(*)\pm}$ momentum spectrum for the
continuum, the on-resonance spectrum with momentum higher than $2.45$ GeV/$c$
and luminosity-scaled off-resonance spectrum are
corrected for efficiency and then fit with the Peterson fragmentation 
function. The momentum spectrum of $D_s^{(*)\pm}$ produced in $B$
decays is obtained by subtracting
the fitted Peterson function from the efficiency-corrected on-resonance spectrum.
The measured branching fractions are:
\begin{figure}
\epsfxsize140pt
\figurebox{120pt}{160pt}{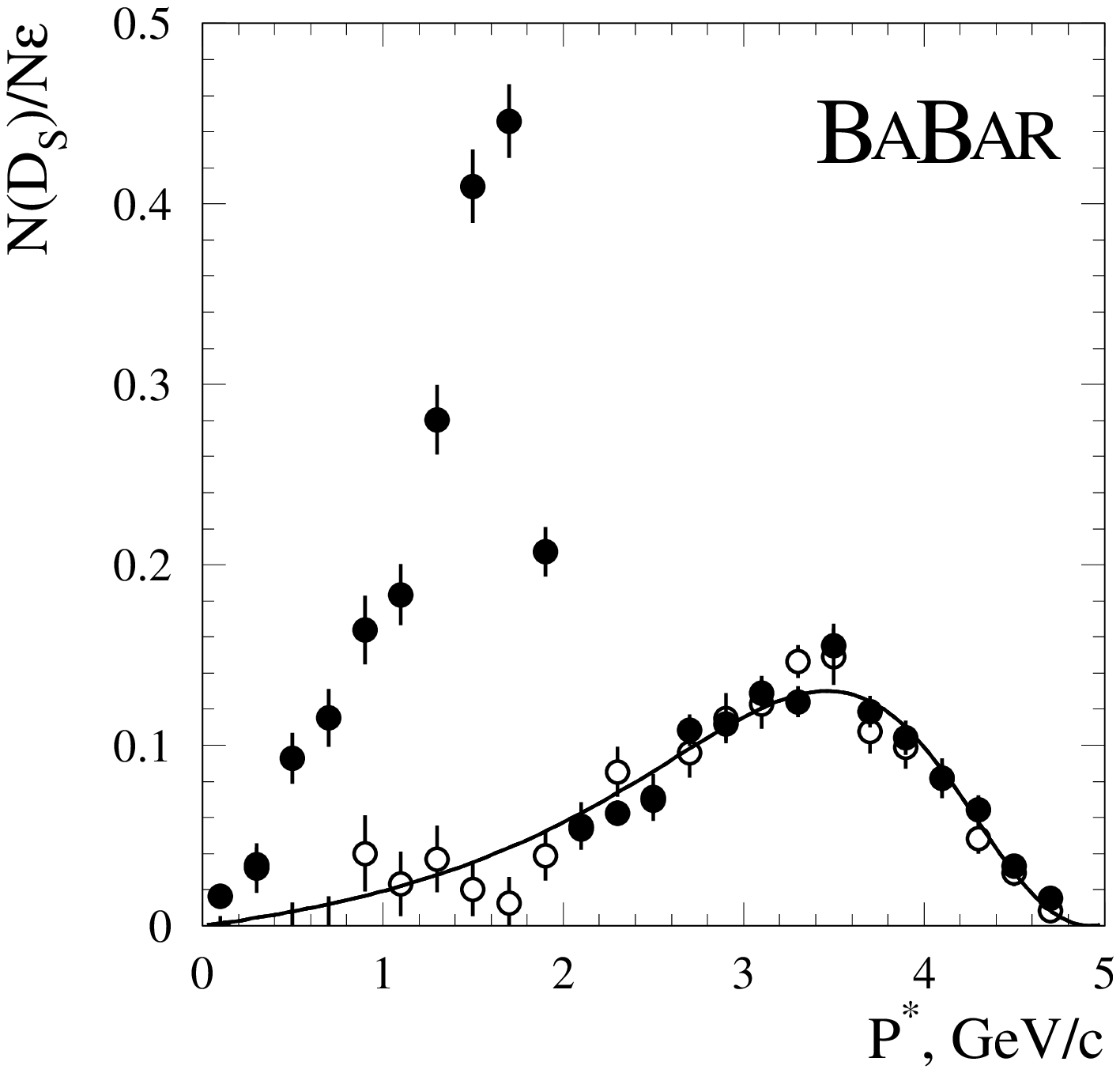}
\epsfxsize140pt
\figurebox{120pt}{160pt}{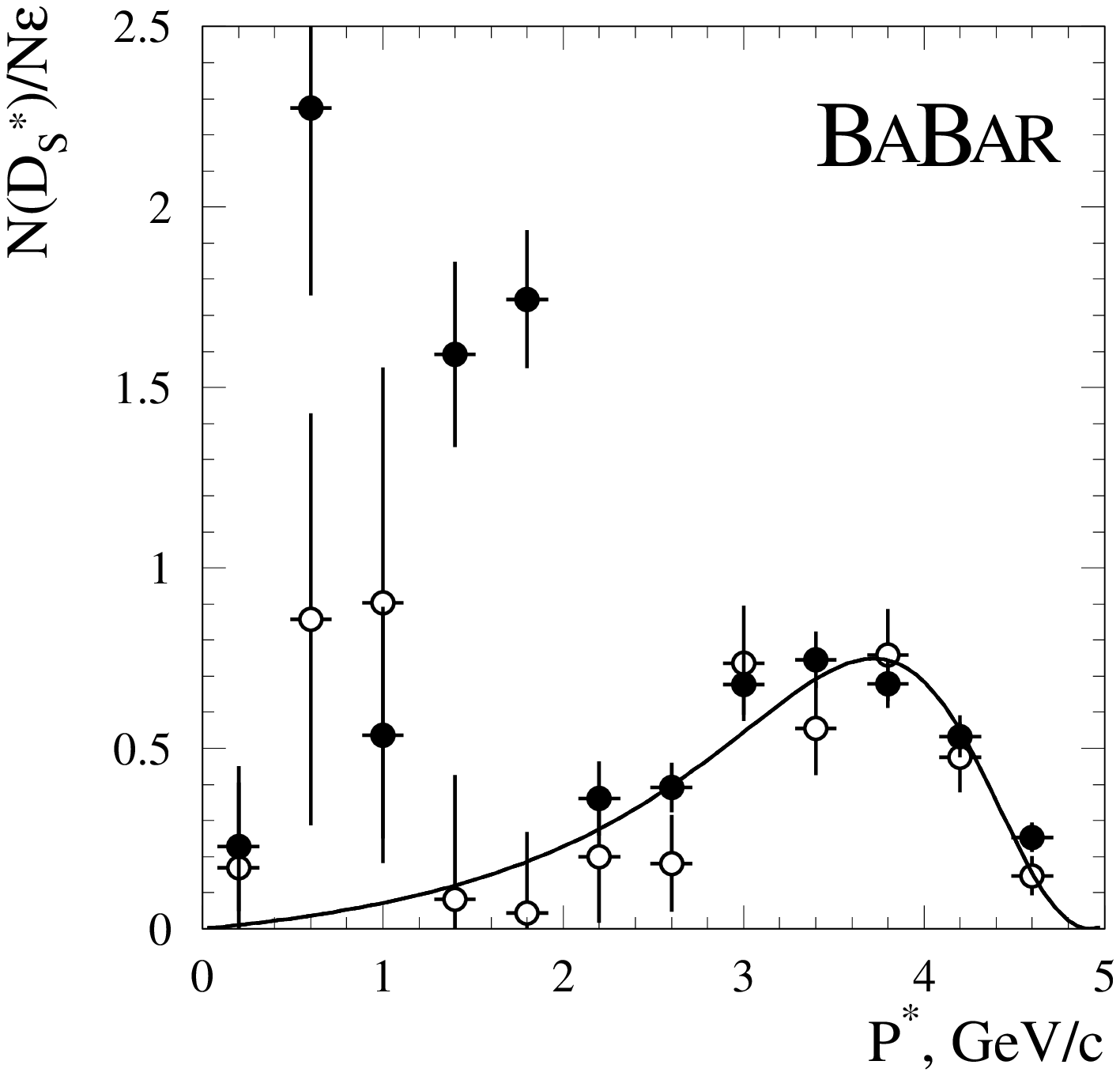}
\vspace*{-0.3cm}
\caption{
Efficiency-corrected momentum spectra for $D_s^{\pm}$ (right)
and $D_s^{*\pm}$ (left). In both figures solid circles represent the on-resonance 
data, while open circles represent data collected off
resonance scaled by luminosity. The solid line shows the
fit to the  Peterson
fragmentation function used for subtracting
the continuum component of the spectrum. }
\label{fig:momspectra}
\end{figure}
\begin{eqnarray}
{\mathcal{B}}(B\rightarrow D_s^{\pm} X) &=& 
\Biggl[(11.9\pm0.3\pm1.1)\times 
\frac{3.6\pm0.9\%}{{\mathcal{B}}(D_s^{\pm}\rightarrow
  \phi\pi^{\pm})}\Biggr]\% 
\label{eq:brdsx}
\end{eqnarray}
\begin{eqnarray}
{\mathcal{B}}(B\rightarrow D_s^{*\pm} X) &=& 
\Biggl[(6.8\pm0.8\pm1.7)\times 
\frac{3.6\pm0.9\%}{{\mathcal{B}}(D_s^{\pm}\rightarrow
  \phi\pi^{\pm})}\Biggr]\% 
\label{eq:brdstx}
\end{eqnarray}
where the first error is statistical, the second 
systematic and the third is due to the 
$D_s^{\pm}\rightarrow \phi\pi^{\pm}$ branching fraction uncertainty.

\section{Measurement of $\Bz \rightarrow D^{*-}D_s^{(*)+}$ Branching Fractions}
\noindent

The measurement of the branching fractions for the decays $\Bz \rightarrow
D^{*-}D_s^{+}$ and $\Bz \rightarrow D^{*-}D_s^{*+}$  uses
a partial reconstruction technique. The $D_s^{(*)+}$ are fully reconstructed, 
we do not reconstruct the $\Dzb$ coming from the
$D^{*-}$  decay.
Instead, we combine a $D_s^{(*)+}$ candidate with a $\pi^-$ and 
then calculate missing mass assuming that the $D_s^{(*)+}$ and $\pi^-$ originate
from the same \Bz.  This missing mass should be the $D^0$ mass if our
hypothesis is correct.
The yield of $\Bz \rightarrow D^{*-}D_s^{(*)+}$ is evaluated by fitting
the missing mass distribution (Fig.~\ref{fig:brdsdst}) with the sum of a 
Gaussian and a background function\cite{conf0013}.
\begin{figure}
\epsfxsize160pt
\figurebox{120pt}{160pt}{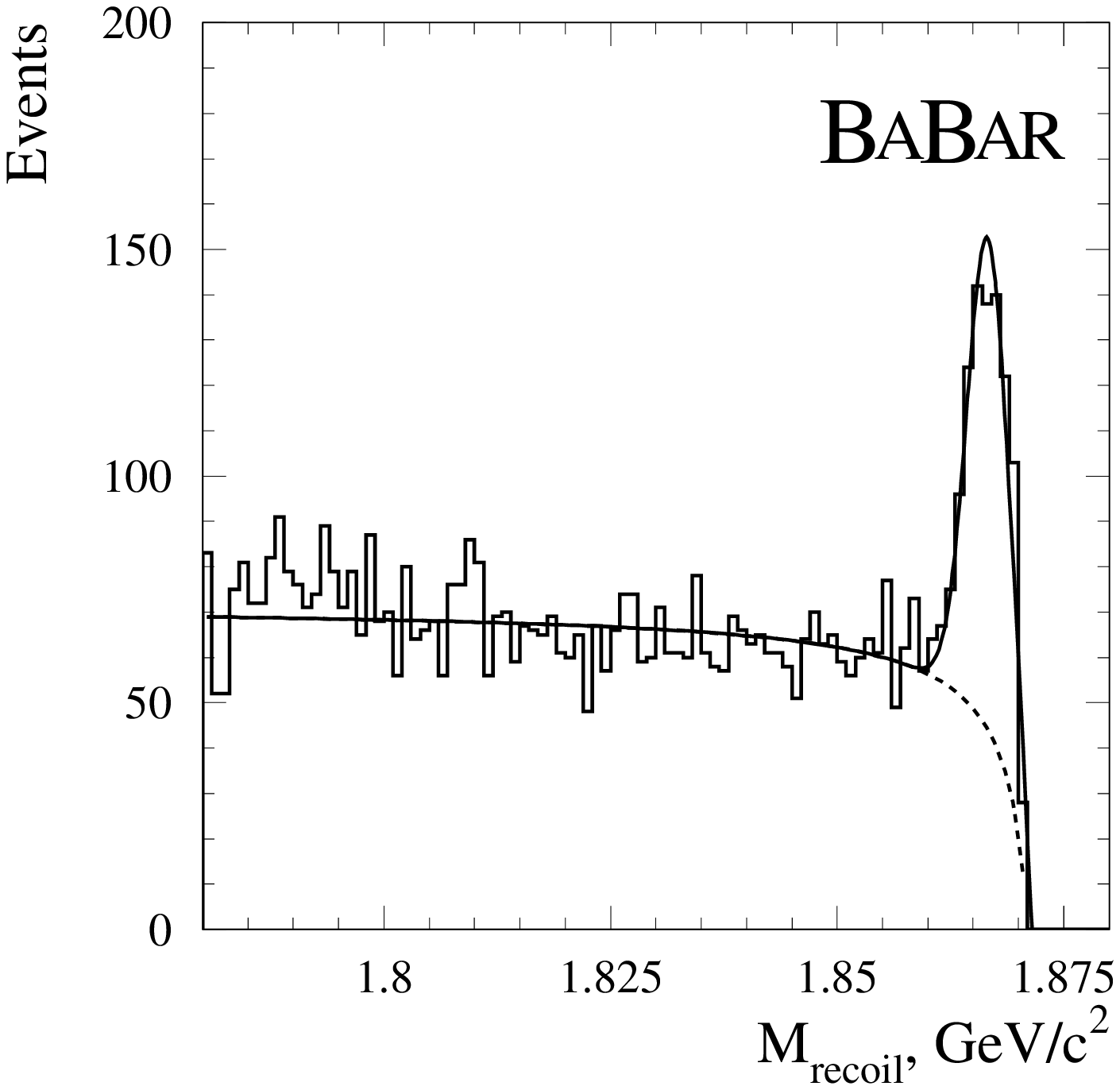}
\epsfxsize160pt
\figurebox{120pt}{170pt}{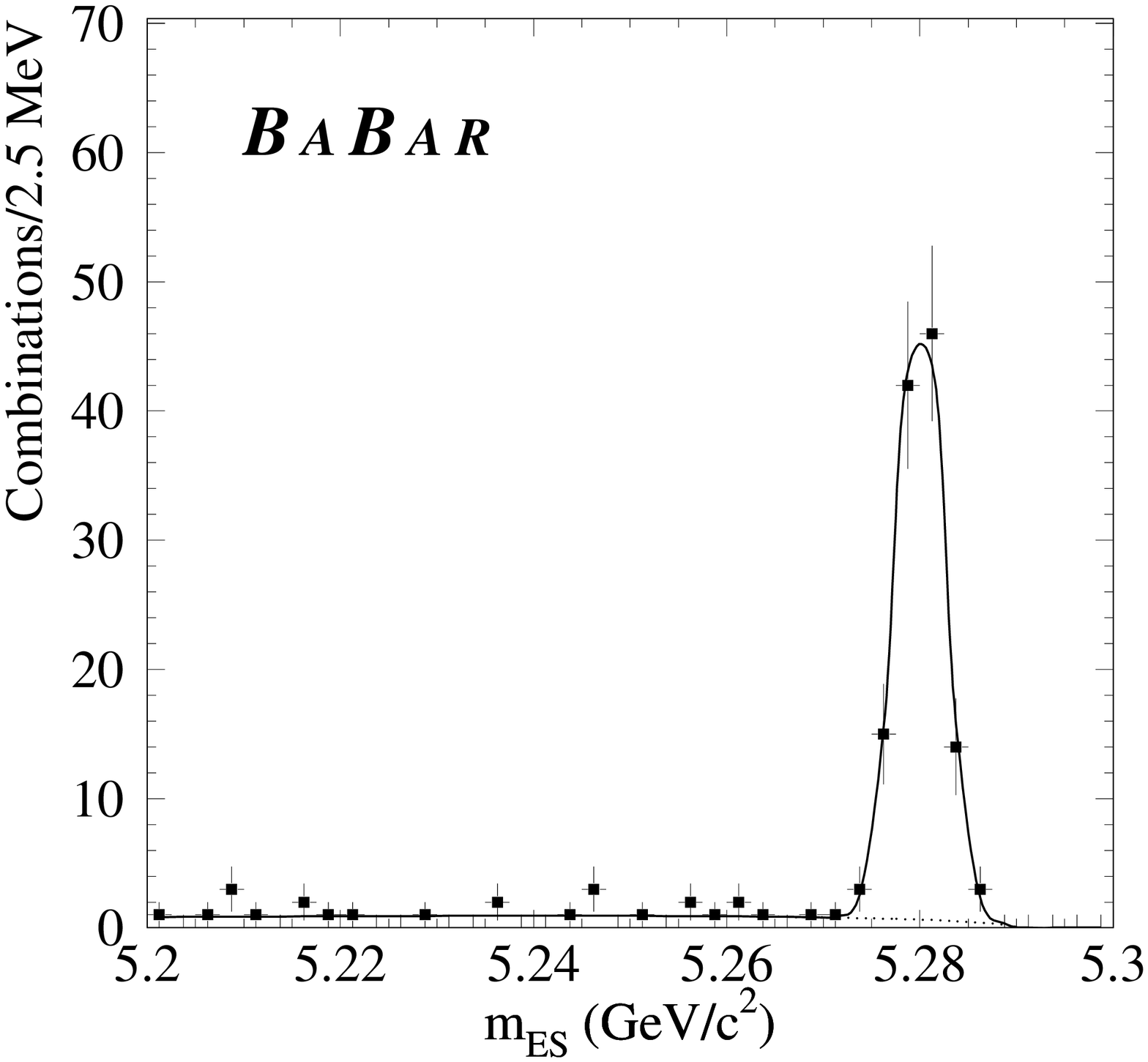}
\caption{Left: Missing mass distributions for the $D_s^{\pm}-\pi$ systems 
before background subtraction.
Right: Distribution of $m_{ES}$ for $|\Delta
 E  |< 2.5\sigma_{\Delta E}$ for the channel 
 $\Bz \rightarrow D^{*-}\pi^+$.  }
\label{fig:brdsdst}
\end{figure}
The branching fractions we measure are $(7.1\pm2.4\pm2.5\pm1.8)\times
10^{-3}$ for the channel  $\Bz \rightarrow D^{*-}D_s^{+}$, and
$(2.5\pm0.4\pm0.5\pm0.6)\times 10^{-2}$ for the channel
for $\Bz \rightarrow D^{*-}D_s^{*+}$.


\section{Measurement of $\Bz \rightarrow D^{*-}\pi+$ and $\Bz
\rightarrow D^{*-}\rho^+$ Branching Fractions} 
\noindent
 
$B^0$ candidates in the channels $D^{*-}\pi^+$ and $D^{*-}\rho^+$ are 
fully reconstructed in the decay chain 
$D^{*-} \rightarrow \Dzb\pi^-$, followed by $\Dzb \rightarrow
K^+ \pi^-$. The $\rho^+$ is reconstructed in the mode $\rho^+\rightarrow \pi^+\pi^0$.
Event selection is based on a few simple criteria. Tracks are
required to originate from near the beam interaction point and
no particle identification is used. Photons with energy grater 
than $30$ \mevcc\ are combined to form $\pi^0$ candidates. To 
form a \Dz\ candidate, kaons and pions 
with opposite charge and originating from the same vertex
must have an invariant mass within $\pm 2.5\sigma$ of the nominal $D^0$ mass.
The $D^0$ candidates are required to have a momentum grater than
$1.3$ GeV/$c$ in the $\upsbb$ frame and are combined with a pion to
form a charged $D^{*}$ candidate.
\Dstar\ candidates are required to have 
$\Delta m = m(\Dzb\pi^-)-m(\Dzb)$ within $2.5\sigma$ of 
the nominal mass difference.
Finally, we combine \Dstar candidates 
with a $\pi^+$, with a momentum grater than $500$ \mevc\ or a $\rho^+$ candidate to form 
$B^0$ candidates.
 In the decay $\Bz \rightarrow D^{*-}\pi^+$ the longitudinal polarization 
 of the  $D^{*-}$ is used to reduce background\cite{conf0006}.
 For the $\Bz \rightarrow D^{*-}\rho^+$ mode, $\rho^+$ candidates are
 selected requiring the $\pi^+\pi^0$ invariant mass  within
 $150$ \mev\ of the $\rho^+$ nominal mass. Event shape variables are also
 used to remove continuum background.
For correctly reconstructed $B$ mesons, the energy of the $B$ candidate, 
$E^*_{B^0}$ 
evaluated in the 
$\upsbb$ frame must be equal to $\sqrt{s}/2$.
We define $\Delta E =E^*_{B^0}-\sqrt{s}/2 $. 
The beam energy substituted mass, $m_{ES}$ is defined as
$  m_{ES}^2  =  \left(\sqrt{s}/2\right)^2 -\left(\sum_i \mbox{\boldmath 
$p$}_i\right)^2$, where  \mbox{\boldmath$p$}$_i$ is the
 momentum of the $i$th daughter of the $B$ candidate.
  The variables $\Delta E$ and $m_{ES}$ are used to define the signal
  and sideband regions.
 For both modes, the region between $5.2$ and $5.3$ GeV/$c^2$  in
 $m_{ES}$ and between  $\pm300$ \mevcc\ in $\Delta E$ is used to study
 signal and background properties. 
 We discriminate against correlated background from $B$ decays where a real
 final state pion (e.g. from higher-multiplicity \Bz\ decays)
 is either not included in the reconstruction, or a
 random one is added to the observed state, with a requirement on $|\Delta E|$.
 The measurement of branching fractions requires an estimate of the number
 of signal events.
A Gaussian and a background function\cite{ref:ARGUS}, which
parametrizes how the phase space approaches zero as the energy approaches
$\sqrt{s}/2$, are used to fit the $m_{ES}$ distribution obtained by requiring
$|\Delta E  |< 2.5\sigma_{\Delta E}$ as shown in
 Fig.~\ref{fig:brdsdst}.

Based on the fitted yield of signal events
the preliminary results for the branching fractions for 
$\Bz \rightarrow D^{*-}\pi^+$ and $\Bz \rightarrow D^{*-}\rho^+$
are $(2.9\pm 0.3\pm 0.3)\times 10^{-3}$ and
$(11.2\pm 1.1\pm 2.5)\times 10^{-3}$, respectively.
The branching fraction for $\Bz \rightarrow D^{*-}\rho^+$ includes all 
non-resonant and quasi-two-body contributions that lead to a $\pi^+
 \pi^0$ invariant mass in the $\rho$ band; however,
the acceptance for non-resonant $D^{*-}\pi^+\pi^0$ decays
is about 15\% 
of $D^{*+}\rho^+$. Moreover, previous measurements
indicate that the
non-resonant branching fraction is comparatively small.
Therefore we estimate
that the non-resonant contribution to our result
is small.
Both branching fraction results compare well with previous measurements
and with the world average\cite{ref:pdg98}.

\textheight=7.8truein
\setcounter{footnote}{0}
\renewcommand{\thefootnote}{\alph{footnote}}

\nonumsection{References}

\end{document}